\documentclass[pre,floatfix,superscriptaddress,notitlepage]{revtex4}
\usepackage[latin9]{inputenc}
\usepackage{graphicx,color}
\usepackage{xspace,subfigure,epsfig}
\usepackage{graphicx,times,latexsym,amsfonts,amssymb,amsmath}
\usepackage{comment,url}
\usepackage{lipsum}

\usepackage{bbm}\usepackage{mathrsfs}\usepackage{float}
\usepackage{cases}

\newcommand{\1}{\mathbbm 1}

\newcommand{\cU}{\mathcal U}
\newcommand{\cS}{\mathcal S}
\newcommand{\cE}{\mathcal E}
\newcommand{\bx}{\mathbf x}

\newcommand{\bz}{\mathbf z}
\newcommand{\bt}{\mathbf t}

\newcommand{\by}{\mathbf y}

\newcommand{\E}{\mathbbm E}
\newcommand{\p}{\mathbbm P}
\newcommand{\dd}{\partial}
\newcommand{\sm}{\setminus}

\newcommand{\ls}[1]
   {\dimen0=\fontdimen6\the\font
    \lineskip=#1\dimen0
    \advance\lineskip.5\fontdimen5\the\font
    \advance\lineskip-\dimen0
    \lineskiplimit=.9\lineskip
    \baselineskip=\lineskip
    \advance\baselineskip\dimen0
    \normallineskip\lineskip
    \normallineskiplimit\lineskiplimit
    \normalbaselineskip\baselineskip
    \ignorespaces
}

\begin{document}

\title{Stochastic Optimization of Service Provision with Selfish Users}

\author{F.~Altarelli}
\affiliation{DISAT and Center for Computational Sciences, Politecnico di Torino, Corso Duca degli Abruzzi 24, 10129 Torino, Italy}
\affiliation{Collegio Carlo Alberto, Via Real Collegio 30, 10024 Moncalieri, Italy}
\author{A.~Braunstein}
\affiliation{DISAT and Center for Computational Sciences, Politecnico di Torino, Corso Duca degli Abruzzi 24, 10129 Torino, Italy}
\affiliation{Human Genetics Foundation, Via Nizza 52, 10126 Torino, Italy}
\affiliation{Collegio Carlo Alberto, Via Real Collegio 30, 10024 Moncalieri, Italy}
\author{C.F.~Chiasserini}
\affiliation{DET, Politecnico di Torino, Corso Duca degli Abruzzi 24, 10129 Torino, Italy}
\author{L.~Dall'Asta}
\affiliation{DISAT and Center for Computational Sciences, Politecnico di Torino, Corso Duca degli Abruzzi 24, 10129 Torino, Italy}
\affiliation{Collegio Carlo Alberto, Via Real Collegio 30, 10024 Moncalieri, Italy}
\author{P.~Giaccone}
\affiliation{DET, Politecnico di Torino, Corso Duca degli Abruzzi 24, 10129 Torino, Italy}
\author{E.~Leonardi}
\affiliation{DET, Politecnico di Torino, Corso Duca degli Abruzzi 24, 10129 Torino, Italy}
\author{R.~Zecchina}
\affiliation{DISAT and Center for Computational Sciences, Politecnico di Torino, Corso Duca degli Abruzzi 24, 10129 Torino, Italy}
\affiliation{Human Genetics Foundation, Via Nizza 52, 10126 Torino, Italy}
\affiliation{Collegio Carlo Alberto, Via Real Collegio 30, 10024 Moncalieri, Italy}

%

\begin{abstract}
We develop a computationally efficient technique to solve a fairly
general distributed service provision problem with selfish users and
imperfect information.  In particular, in a context in which the
service capacity of the existing infrastructure can be partially
adapted to the user load by activating just some of the service units,
we aim at finding the configuration of active service units that
achieves the best trade-off between maintenance (e.g.\ energetic)
costs for the provider and user satisfaction.  The core of our
technique resides in the implementation of a belief-propagation (BP)
algorithm to evaluate the cost configurations.  Numerical results
confirm the effectiveness of our approach.
\end{abstract}

\maketitle

\section{Introduction}

Mathematical models of the distributed service provision problem have been studied thoroughly in computer science under the name of {\em selfish load balancing} \cite{V07} and {\em congestion games} \cite{R73,R73b}.
Most results apply concepts borrowed from Game Theory and concern worst-case analysis, in particular the  computation of the so-called ``price of anarchy", i.e. the ratio between
the cost of the worst Nash Equilibrium (NE) and the optimal social cost \cite{KP99}. Several works also address algorithmic issues, such as the question of designing distributed dynamics that converge to NEs, their convergence time \cite{EDKM03}, or computational complexity \cite{FKKMS02}.
In many practical problems, service providers should be more interested in the average-case scenario, in particular in the average cost of service/resource allocations determined by the selfish behavior of users.
In order to be able to compare the expected cost of different service
allocations, a service provider is called to the arduous computational
task of evaluating an average over the possibly huge number of
different NEs that are obtained as result of the allocation. In
addition, service providers do not always have perfect information
about the user behavior -- a fact that is usually modeled by including some stochastic parameter into the problem \cite{GI99,KRT97,NSM11,DST03}.
In the presence of stochasticity, algorithms based on Monte Carlo methods are extremely inefficient even for moderately large problem sizes, whereas recent works \cite{ABRZ11} have shown that much better results can be obtained using message-passing algorithms inspired by statistical physics methods.

In our formulation,  we assume that the total service capacity of the existing infrastructure can be partially adapted by activating or deactivating some of the service units. Our goal is to find the  configuration of active server units that achieves the best trade-off between maintenance costs  for the provider and user satisfaction. For the sake of example, we assume maintenance costs expressed in terms of energy costs to keep the service unit active.
For any given configuration of service units and users, we propose to use a belief-propagation (BP) algorithm to evaluate the  cost of every service  configuration. Moreover, we put forward an approximate method, also based on BP, which allows to perform the average over the stochastic parameters within the same message-passing algorithm used to average over the NEs.  The information obtained is then used to optimize the service units allocation. This can be done easily either exhaustively or by means of decimation methods.

\section{System Model\label{sec:sys-model}}

The service provision problem is represented
by a bipartite graph $G=(\cU,\cS;\cE)$, in which $\cU=\{1,\dots,U\}$
and $\cS=\{1,\dots,S\}$ are the sets of nodes, {\em users} and {\em service
units} respectively, and $\cE$ is the set of edges $(u,s)$
connecting nodes $u\in\cU$ and $s\in\cS$. In general the graph
is not complete, i.e.,  users cannot connect to any service
unit. In addition, every service unit $s$ has an
operational energy
cost, $r_s$. Thus, in some time periods it may be convenient to keep  active only part of
the
existing service units ($x_{s}=1$) and deactivate the others  ($x_{s}=0$).

The first ingredient of the model is the rational behavior of the
users. In many problems, such as selfish load balancing \cite{V07},
this is introduced by assuming that
the quality of service received by a user, when selecting a service
unit, depends on the load of the unit at the time of service,
defining a correlation between users' utilities. Here, we simplify the
model assuming that
users' satisfaction in selecting a service unit does not
depend on the state of the service unit itself (provided it is available).
Each edge $(u,s)$ has a weight, $w_{us}\in\mathbb{R}$, that represents
the satisfaction obtained by  user $u$ when selecting service unit $s$.
However, users' decisions are not independent, as there is a limitation to the number of individuals
 that can be served at the same time by the same service unit.
The weight in the opposite direction, $w_{su}\in\mathbb{R}^{+}$, is the workload sustained by
the service unit $s$ when providing the service
to user $u$. If we assume that each service unit $s$ can sustain a maximum load
 $c_{s}$, the sum of the workloads $w_{su}$ of all users $u$ selecting
unit $s$ should not exceed $c_{s}$. This set of hard \emph{capacity
constraints} introduces an indirect competition between users.
More precisely, suppose that user $u$ considers service unit $s$ to
be the preferable one (i.e.,
$w_{us}\geq w_{us'}$ $\forall s'\neq s$),
but adding the workload $w_{su}$ of $u$ to the total load already
faced by unit $s$, it would exceed $c_{s}$. Then we say that service
unit $s$ is \emph{saturated} for user $u$ and the latter has to
access another of the service units accessible to her. She will thus
turn to the unit with the second best weight. If this one is \emph{available},
user $u$ will make use of it, otherwise she will try the third one
on her list and so on. Note that multiple connections from the same
user to many service units are not allowed.

The second ingredient is stochasticity. We imagine that in any realistic situation the activity of the users
could follow very complex temporal patterns. Users could leave the
system and come back, using different service units depending on their
preference and the current availability.
The stochastic nature of the problem is summarized into a set
of stochastic parameters $\{t_{u}\}_{u\in\cU}$.
At any given time, with probability $P\left(t_{u}=0\right)$  the user $u$ is absent from the service
system and $t_{u}=0$,  whereas  with probability $P\left(t_{u}=1\right)$ she is present
and $t_{u}=1$.
For the moment, we can assume that the actual realization
$t_{u}$, $\forall u\in\cU$, is known.

Given the bipartite graph, the configuration of active service units $\{x_{s}\}_{s\in\cS}$
and the set of parameters $\{t_{u}\}_{u\in\cU}$, every user tries
to maximize her own utility using the best service unit available
(i.e., among those that are not saturated or inactive).
Such a  system model  can represent several different
application scenarios. For example, we can represent   videoconferencing, including several Multipoint
Control Units (MCUs) or a heterogeneous wireless access network,
where  points of access, possibly  using  different technologies, are available (e.g., 3G/LTE, WiFi,
WiMax) to the users. In these scenarios, indeed, it would be convenient to switch off service units when underloaded.

\section{Problem Definition\label{sec:problem}}

The system outlined above poses the following service provision
problem:
{\em at any given time
period, which service
units should  be activated by a central controller, in order to maximize the users' satisfaction and
minimize the system energy consumption?}
Since the decision of the central controller
has to account for the rational behavior of the users, we address
the optimization problem as follows.

First, we consider a system configuration, where the active service
units are given, and model the users' association process as a game.
The players of the game are the users that have to
select a service unit among the active ones.
We solve the game so that, for each user pattern,
$\{t_{u}\}_{u\in\cU}$, the corresponding NE strategy profiles can be identified;
note that, given $\{t_{u}\}_{u\in\cU}$, there may exist multiple
NEs. Then, in order to evaluate the performance of the system configuration,
we define an objective function, which accounts for the energy cost of
the active units and for the users' satisfaction.
Since, for a given $\{t_{u}\}_{u\in\cU}$, different NEs are reached depending
on the order of arrival of the users, we average the objective
function first over all NEs corresponding to $\{t_{u}\}_{u\in\cU}$,
and then
over all possible realizations of the users arrival process.
Finally, we use the obtained result as an indication of
the system configuration performance, and we select the system
configuration that optimizes such an index.

Let us now detail the procedure outlined  above.
We denote the tagged system configuration by $\bx$, and define $\cS_u$
as the set of service units that can be selected by user
$u \in \cU$. Also, let $\cU_s$ be the set of users that can select service unit
$s \in \cS$ and let $\cU_{s,u}$ be the set of nodes $v\in\cU_s\sm u$
with $s\in\cS_u$.

In the game, the action of the generic user (player) $u$ consists in choosing one of the service units
connected to her, e.g., $z_{u}=s$ with $s\in\cS_u$. The payoff
is $w_{us}\geq0$ if unit $s$ is active and not saturated otherwise
it is $-\infty$. If no unit is chosen, $z_{u}=\emptyset$ and the
payoff is $-\omega$, being $\omega$ a penalty value. More precisely:
\vspace{-3mm}
\begin{align}
\pi_{u}(z_{u},\{z_{v}\}_{v\in\cU_{s,u}}|t_u=1) &=
\begin{cases}
-\omega, & \text{if }z_{u}=\emptyset\\
w_{us}, & \text{if }z_{u}=s,\text{ and }\\
& w_{su}+\sum_{v\in\cU_{s,u}}\delta(z_{v},s)w_{sv}\leq c_{s}x_{s}\\
-\infty, & \text{if }z_{u}=s,\text{ and }\\ &w_{su}+\sum_{v\in\cU_{s,u}}\delta(z_{v},s)w_{sv}>c_{s}x_{s}
\end{cases}
\label{eq:pi}
\end{align}
If instead user $u$ is absent, $z_u=\emptyset$ is the only possible value and we set
\vspace{-3mm}
\begin{align}
\pi_{u}(z_{u},\{z_{v}\}_{v\in\cU_{s,u}}|t_u=0)=\begin{cases}
0, & \text{if }z_{u}=\emptyset\\
-\infty, & \text{otherwise.}
\end{cases}
\nonumber
\end{align}
Note that, at every perturbation in the system, e.g. due to the
departure of a user, a player may decide to connect to
another service unit than the currently selected one, if she can
increase her payoff.

It is useful to represent the NE
conditions in terms of \emph{best-response relations}: a strategy
profile $\bz^{*}=(z_{1}^{*},\dots,z_{N}^{*})$ is a pure NE
if and only if, for each user $u$, $z_{u}^{*}$ is the best response
to the actions of the others, i.e.,
\begin{align}
z_{u}^{*}=\arg\max_{z_{u}}\pi_{u}(z_{u},\{z_{v}^{*}\}_{v\in\cU_{s,u}}| t_u),\quad\forall u\in\cU \,.
\end{align}
In principle, the weight given to each NE should depend
on specific details of the dynamics of the users (e.g. on the order of arrival of the users and on the order according to which users
unilaterally deviate from the current strategy profile if they find it convenient). Unfortunately these details are largely unknown in any realistic setup.
It is thus worth considering a simplifying hypothesis in which all the NEs are weighted uniformly and the complex user dynamics is summarized in the
average over the realizations of the stochastic parameters $\bt$. In general we do not know which users are actually present in the
system, but we assume to know the probability $p_{u}$
that user $u$ is present, $\forall u\in\cU$.

The problem consists in optimizing the trade-off between
the system energy cost and the expected users' satisfaction, i.e. in finding the configuration
of active service units $\{x_{s}\}$ which maximizes the following
objective  function:
\begin{align}
\mathcal{F}(\bx) = \E_{\bt}\left[ \left\langle
  \sum_{u\in\cU}\pi_{u}(z_{u},\{z_{v}\}_{v\in\cU_{s,u}}|t_{u}) \right\rangle \right] - \alpha\sum_{s}r_{s}x_{s}
  \nonumber
\end{align}
where $\left\langle \cdots \right\rangle$ represents the average over the values of $\bz \in \textsc{Nash}(\bx,\bt)$ that satisfy the NE conditions (which depend implicitly on $\bx$ and $\bt$). The objective function is composed of two contrasting terms: a first contribution that measures the overall quality of the service, and a second term that quantifies the
total cost of active service units (alternatively, the service provider's
revenue could depend explicitly on the perceived quality of the service).
The parameter $\alpha$ is used to set the relative importance of
the two objectives.

We can finally formulate the optimization goal
of the central controller which is, given $G=(\cU,\cS,\cE)$, the vector
of unit capacities $\mathbf{c}$, the payoff matrix $\mathbf{W}$,
the vector of presence probabilities $\mathbf{t}$, and the parameter
$\alpha$, to find a minimizing $\bx^{*}$ such that
\(
\mathcal{F}\left(\bx^{*}\right)=\min_{\bx}\mathcal{F}\left(\bx\right)
\).
In conclusion, the vector $\bx^{*}$ represents the system configuration
that corresponds to the best tradeoff between the system energy cost
and the user satisfaction.

\section{Problem Solution\label{sec:solution}}

The NE conditions define a set of local hard constraints
on the individual actions, such that finding a pure NE can be translated
into finding a solution of a Constraint-Satisfaction Problem (CSP).
Using the node variables $\{z_{u}\}_{u\in\cU}$,  we can formulate  the
associate CSP  over  a factor graph with many small loops even when the original graph
had none, which is not appropriate to develop a solution algorithm
based on message passing \cite{MM09}. In the following we switch to an equivalent
representation, using edge variables, that is much more convenient
for message passing applications.

\subsection{CSP Representation Using Edge Variables\label{subsec:NEs}}

The actions of the users can be described using three-states variables
$y_{us}$ defined on the undirected edges $(u,s)\in E$ (see Figure
\ref{fig2})
\begin{align}
y_{us}=\begin{cases}
-1, & \text{if }s\text{ is inactive or saturated for }u\\
0, & \text{if }s\text{ is available but not used by }u\\
1, & \text{if }s\text{ is used by }u
\end{cases}
\end{align}
where ``saturated for $u$'' refers to the case in which  if $u$ connects
to $s$, the latter violates the capacity constraint, while
``available'' refers to the case where $s$ is active and able to accommodate user $u$.
The NEs are the configurations taken by the variables
$\{y_{us}\}$ that satisfy the following set of constraints.
First, users cannot have
access to more than one service unit at the same time:  $\sum_{s\in\cS_u}\1\left[y_{us}=1\right]\leq t_{u} \quad (\forall u\in\cU)$. Second,
the capacity of each service unit cannot be exceeded: $\sum_{u} \1\left[ y_{us} = 1\right] w_{su} \leq c_s x_s \quad (\forall u\in\cU_s, s\in\cS)$. And third, users try to use the best service unit available: $\{y_{us}=1, y_{us'}=0\} \Rightarrow \{w_{us} \geq w_{us'}\} \quad \forall (s,s'\in\cS_u, u\in\cU)$.

The stochastic optimization problem consists in finding
the configuration of active service units $\bx$ such that it maximizes the objective function
\vspace{-2mm}
\begin{equation}
\mathcal{F}(\bx)= \E_{\bt}\left[\left\langle
    \sum_{(u,s)\in \mathcal E} {\1}_{[y_{us}=1]}w_{us} \right\rangle \right] -\alpha\sum_{s}r_{s}x_{s}. 
\end{equation}
The most difficult part of this optimization problem is that of performing
the average over the NEs in the presence of stochasticity,
that is the essential step to be able to evaluate the average costs
and benefits from activating/deactivating different service units. Once
this is done, the optimization step over the $\{x_{s}\}$ becomes
trivial and it can be done either exhaustively or by means of decimation
methods. In the next section we describe an approximate method to
perform the average over the NEs and the stochastic parameters
in a computationally efficient way.

\subsection{Average over NEs with Stochastic Parameters\label{subsec:ave-NE}}

In general one should first average over the pure NEs at fixed realization $\bt=\{t_{u}\}$ of the stochastic parameters and then perform the average over the distribution $P(\bt)=\prod_{i}P\left(t_{i}\right)$ of the latter.
The double average is extremely costly at a computational level. The message passing approach allows one to perform these two steps together although at the cost of introducing an approximation in the computation.

\begin{figure}
\centering
\includegraphics[width=10cm]{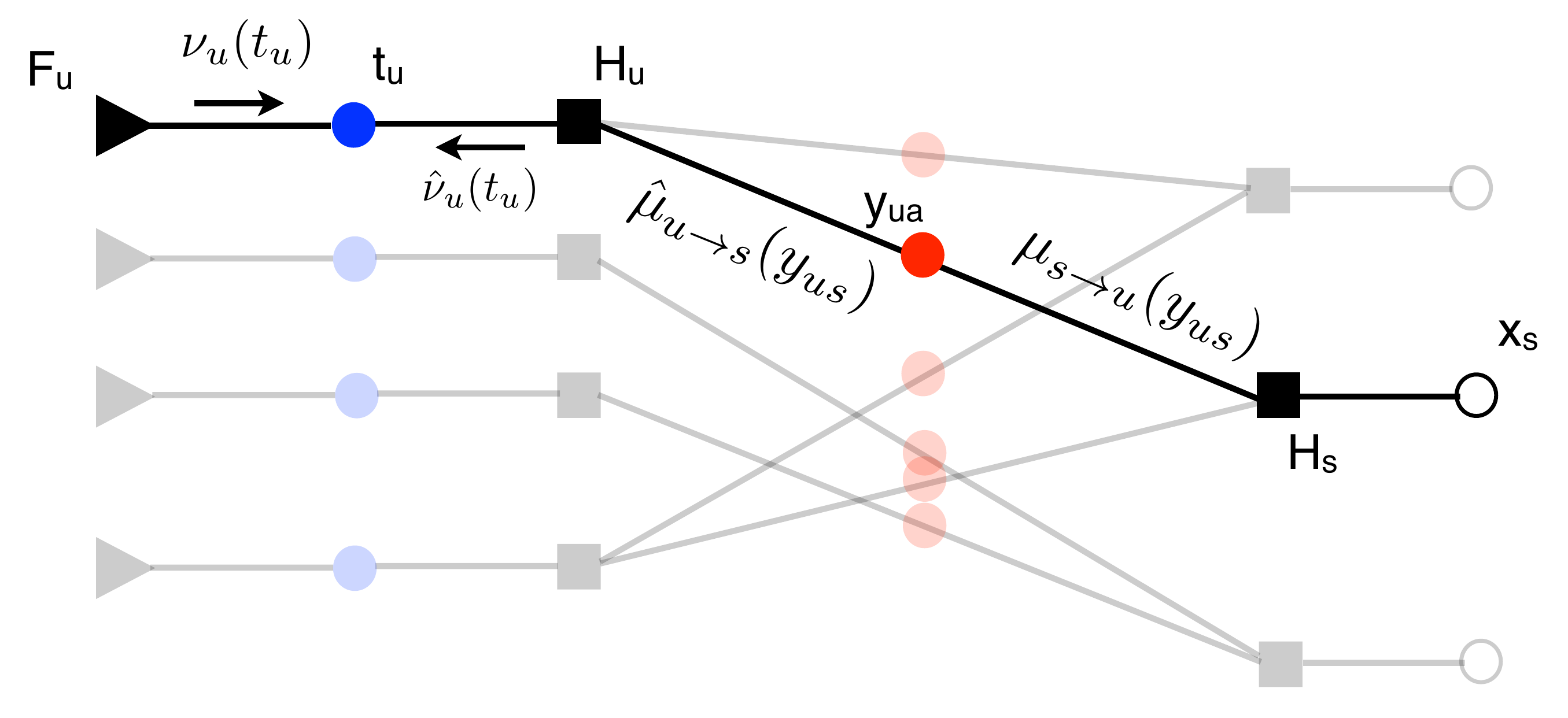}
\caption{Factor graph representation of Eqs. \eqref{bp1}-\eqref{bp4} involving the four types of messages. The use of edge variables $\{ y_{us}\}$ highly simplifies the representation and allows to get rid of small loops. The other two types of variable nodes are blue and empty nodes, corresponding respectively to  the stochastic parameters $\{ t_u\}$ and the service provider's variables $\{x_s\}$. Square nodes $\{H_u\}$ and $\{H_s\}$ are standard factor nodes containing the capacity constraints and the best-response conditions.}
\label{fig2}
\end{figure}

For an observable $O(\by,\bx,\bt)$, the average over NEs is
\begin{eqnarray}\label{quenched}
\mathcal{O}(\bx) & = &  \sum_{\by}\sum_{\bt}P(\by,\bt|\bx)O(\by,\bx,\bt)\nonumber\\
 & = & \sum_{\by}\sum_{\bt}P(\bt)P(\by|\bt,\bx)O(\by,\bx,\bt)\nonumber \\
 & = & \sum_{\by} \sum_{\bt}P(\bt)\frac{\1\left[\by\in\textsc{Nash}(\bx,\bt)\right]}{Z(\bx,\bt)} O(\by,\bx,\bt)
\end{eqnarray}
in which $Z(\bx,\bt)=\sum_{\by}\1[\by\in\textsc{Nash}(\bx,\bt)]$.
The numerator can be easily expressed in terms of the local constraints for the edge variables $\{y_{us}\}$. If we call $I_{s}(\{y_{vs}\}_{v\in\cU_s}|x_{s})$ the
hard constraint defined on node $s\in\cS$ and $I_{u}(\{y_{us}\}_{s\in\cS_u}|t_{u})$ that defined on $u\in\cU$, we have
$\1\left[\by\in\textsc{Nash}(\bx,\bt)\right] = \prod_{u}I_{u}(\{y_{us}\}_{s\in\cS_u}|t_{u})\prod_{s}I_{s}(\{y_{vs}\}_{v\in\cU_s}|x_{s})$.
The main difficulty of performing the quenched average is due to the presence of the normalization factor $Z(\bx,\bt)$ at the denominator of \eqref{quenched}.
A mean-field approximation, based on the factorization ansatz $Z\left(\bx,\bt\right)=\prod_{u}Z_{u}\left(\bx,t_{u}\right)$, can be used to transform our quenched average into
an easily computable annealed one. In this approximation we get
\begin{align}
\log{P\left(\by,\bt|\bx\right)}  =  \sum_{u}\log P\left(t_{u}\right) -\sum_{u}\log Z_{u}\left(t_{u}\right)
+ \sum_{u}\log I_{u}\left(\{y_{us}\}_{s\in\cS_u}|t_{u}\right)
+ \sum_{s}\log I_{s}\left(\{y_{us}\}_{u\in\cU_s}|x_{s}\right).
\end{align}

The factor graph associated to the problem is shown in Figure \ref{fig2}. In addition to the usual terms $H_u = \log I_{u}\left(\{y_{us}\}_{s\in\cS_u}|t_{u}\right)$ and $H_s = \log I_{s}\left(\{y_{us}\}_{u\in\cU_s}|x_{s}\right)$, corresponding to hard constraints, it also contains energetic terms $\log Z_{u}\left(\bx,t_{u}\right)$ on the nodes $u\in \mathcal{U}$. The energetic terms are unknown but can be computed implicitly introducing a new set of
messages $\{\nu_{u}\left(t_{u}\right)\}_{u\in\mathcal{U}}$ and $\{\hat{\nu}_{u}\left(t_{u}\right)\}_{u\in\mathcal{U}}$ that must be adjusted in order to have the correct probability marginal $P(t_u)$ on each variable node $t_u$. On such a factor graph, it is possible to derive the following set of message passing equations
\begin{align}\label{bp1}
\hat{\mu}_{u\to s}\left(y_{us}\right) & \propto  \sum_{t_{u}}\nu_{u}(t_{u})\sum_{\{y_{us'}\},s'\in\cS_u\sm s}I_{u}(y_{us},\{y_{us'}\}|t_{u})\prod_{s'\in\cS_u\sm s}\mu_{s'\to u}(y_{us'})\\
\mu_{s\to u}\left(y_{us}\right) & \propto  \sum_{\{y_{vs}\},v\in\cU_{s,u}}I_{s}(y_{us},\{y_{vs}\}|x_{s})\prod_{v\in\cU_{s,u}}\hat{\mu}_{v\to s}(y_{vs})\\
\hat{\nu}_{u}\left(t_{u}\right) & \propto  \sum_{\{y_{us}\},s\in\cS_u}I_{u}\left(\{y_{us},s\in\cS_u\}|t_{u}\right)\prod_{s\in\cS_u}\mu_{s\to u}\left(y_{us}\right)  \\
\nu_{u}\left(t_{u}\right) & \propto  P\left(t_{u}\right)\hat{\nu}_{u}^{-1}\left(t_{u}\right).\label{bp4}
\end{align}
by means of which the observable $\mathcal{O}(\bx)$ can be approximately evaluated. The proportionality symbol means that the marginal probabilities need to be correctly normalized. A complete description of the method will be presented elsewhere \cite{tech-rep}.
\setcounter{equation}{9}

\section{Numerical results on random graphs\label{sec:results}}
In this section, we present some numerical results obtained by our
algorithm on random graphs generated with the following
procedure. Both the users and the service units  are placed at random
in the unit square of the two-dimensional euclidean space. For each
user $u$, only the $k$ nearest service units are assumed to be accessible, and, for each of these, the workload is $w_{su} = \left\lceil \gamma \, d(u,s)^2 \right\rceil$, i.e., an integer proportional to the square of the distance between $s$ and $u$ (the proportionality constant $\gamma$ is such that the maximum weight is equal to a specified value $w_{\text {max}}$). Recall that the payoff  for a disconnected user is  $-\omega$ (see (\ref{eq:pi})), whereas for connected users is $w_{us} = w_{\text {max}} - w_{su}$.   Finally, the presence probabilities $p_u = \p[t_u = 1]$ are extracted uniformly in $(0,1]$.
We have considered four scenarios, whose parameters are reported in Table~\ref{tab:para}; the ranges for all these parameters are such that the instances are non trivial.

\begin{table}[!tb]
\begin{center}
\caption{Scenarios considered in the simulations}\label{tab:para}
\begin{tabular}{|l|c|c|c|}
\hline
Scenario & S1 & S2 & S3\\
\hline
Number of instances & $1842$ & $91$ & $1$\\
Number of users for each instance $N_u$ &$ 12$ & $300$ & $1000$\\
Number of service units $N_s$ & $4,8,12$ &$ 30,60,90$ &$ 50$\\
Connectivity of users $k_u$ & $2,3,4$ & $3,5,8$ & $5$\\
Capacity of service units $c_s$ & $5,8,11$ & $5,10,15$ & $20$\\
Maximum weight  $w_{\rm{max}}$ & $10$ & $10$ & $15$\\
Penalty for a disconnected user $\omega$ & $10$ & $10$ & $5$\\
\hline
\end{tabular}
\end{center}
\end{table}

\subsection{Comparison with exhaustive enumeration}
As a first test of our message passing approach, in scenario S1 we compared it to an exhaustive enumeration of the NEs for fixed $\bt$, averaging the results over a sample of values of $\bt$. More specifically, we considered (for a given configuration of the service units $\bx$) the following two observables, in terms of which one can compute the objective function we propose to use for the greedy procedure:
\begin{eqnarray}
 \mathcal W(\bx) = \E_{\bt} \left[ \left< \sum_{(u,s) \in \mathcal E} {\1}_{[y_{us}=1]}w_{su}\right\rangle \right]
\end{eqnarray}
which is the average (over the realization of $\bt$ and over the NEs) of the sum of the workloads $w_{su}$ for the users that are present and connected to some service unit, and
\begin{eqnarray}
 \mathcal N(\bx) = \E_{\bt} \left[ \left< \sum_{u \in \mathcal U} \prod_{s \in \dd u} {\1}_{[y_{us} = -1]} \right> \right]
\end{eqnarray}
which is the average (again, over the realization of $\bt$ and over the NEs) of the number of disconnected users. We compare the value obtained by our algorithm for $\mathcal W(\bx)$ with
\begin{eqnarray}
 \mathcal {\bar W(\bx)} = \frac 1 {|\mathcal T|} \sum_{\bt \in \mathcal T} W(\bx, \bt)
\end{eqnarray}
where $\mathcal T$ is a random sample of realizations of $\bt$
extracted from $P(\bt)$ (and $|\mathcal T|$ is the size of the sample)
and where $W(\bx, \bt)$ is the average over the NEs (for fixed $\bx$
and $\bt$) of the sum of the workloads for connected users, which is
computed with an exhaustive enumeration of the possible allocations
$\by$. A similar comparison is done for $\mathcal N(\bx)$.
Of course, in this scenario the number of users is limited since the exhaustive enumeration is possible only if the size of the instance is very small.

\begin{figure*}
\centering
\includegraphics[width=7cm]{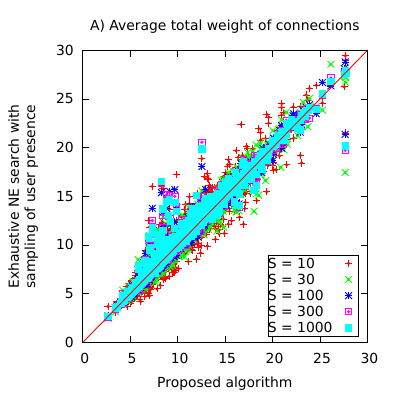}
\includegraphics[width=7cm]{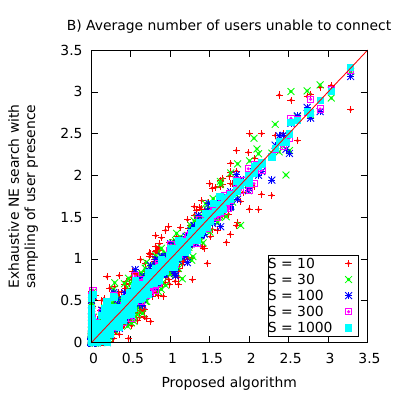}\\
\includegraphics[width=7cm]{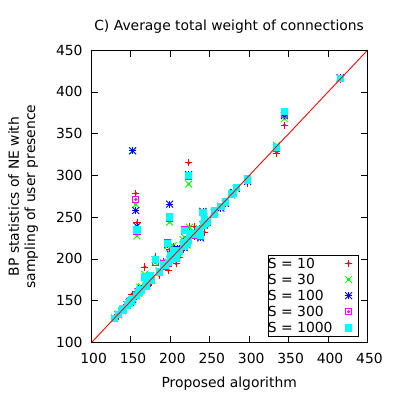}
\includegraphics[width=7cm]{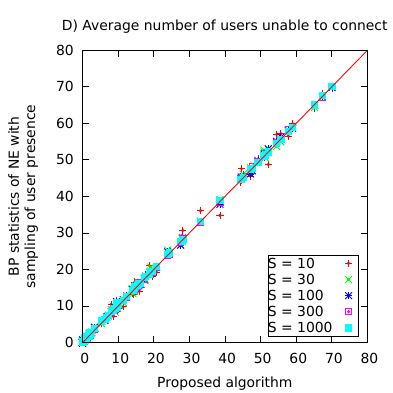}
\caption{Comparison between the values of $\mathcal W(\bx)$ (plots A and C for scenario S1 and S2, respectively) and
  $\mathcal N(\bx)$ (plots B and D for scenarios S1 and S2, respectively), as computed by our message passing
  algorithm (``Mirror'') and by either exhaustive enumeration of the NEs (plots A and B) or
BP sampling of the NEs (for fixed $\bt$) with explicit sampling over $\bt$ (plots C and D).
Each data point
  corresponds to one instance and one
  sample size $S$ (corresponding to the different
  symbols/colors). }
 \label{figComparison}
\end{figure*}

Scatter plots A and B in Fig.~\ref{figComparison} compare  our algorithm with the exhaustive enumeration under scenario S1.
As the sample size $S = |\mathcal T|$ increases, the data
points tend to collapse onto the diagonal, i.e., as the accuracy of the sampling procedure improves, the results obtained by sampling tend to those obtained with the message passing algorithm, except for a small number of ``outliers'' (less than one percent of the instances). This confirms that, even on very small instances, the two hypotheses on which our method is based, namely the decorrelation assumption of the cavity method and the factorization hypothesis for the partition function $Z(\bt)$, are a good approximation.

\subsection{Comparison with explicit sampling}
In the next scenario S2, we compare the results obtained by our algorithm with those
obtained by computing the average over the NEs (for fixed $\bt$) with
BP, and then averaging over $\bt$ with an explicit sampling. This
allows us to test, on larger instances, the factorization assumption
for the partition function $Z(\bt)$. Note that our algorithm requires
only one convergence of the message passing procedure to perform both
averages. The explicit sampling, instead, requires $S$ convergences of
a message passing, which is (almost) as complex as ours; thus, it  is
roughly slower by a factor $S$. This limits the number of instances
that we have been able to analyze to less than 100.

Again, scatter plots C and D of  Fig.~\ref{figComparison} show that, as the sample size $S$ increases, the data points tend to collapse onto the diagonal, with the exception of a few cases for the estimation of $\mathcal W(\bx)$.

\subsection{Optimization results}
Finally, in scenario S3 we provide an example of optimization. We used our greedy decimation heuristic based on the message passing algorithm for a single instance.
The heuristic we use to find the optimal allocation $\bx$ is the following. We start by computing the value of the objective function
\begin{eqnarray}
 \mathcal O(\bx) = \E_{\bt}\left[\left\langle
    \sum_{(u,s)\in \mathcal E} {\1}_{[y_{us}=1]}w_{us} \right\rangle \right]
\end{eqnarray}
when all the service units are on (i.e.\ $x_s = 1$ for each $s$). Then, we compute the same objective function for all the configurations obtained by switching off one service unit. We actually switch off the service unit that corresponds to the smallest drop in the objective function. The same procedure is then iterated, computing the variations in the objective function associated to switching off each of the service units that are still on, and actually switching off the one that minimizes the drop, until all the service units are off (or we decide to stop).

\begin{figure}
 \centering
 \includegraphics[width=10cm]{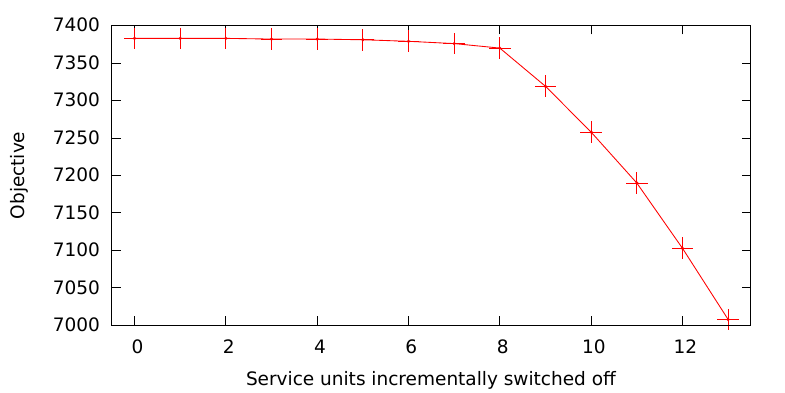}
 \caption{The results of the greedy decimation. The labels on the abscissa show the number of service units being switched off at each step of the decimation. The last five steps of the decimation are discarded from the solution.}
 \label{fig4}
\end{figure}

The results of this ``greedy decimation'' are shown in Fig.~\ref{fig4}. We observe that during the first 8 steps of the decimation (i.e. as we switch off the first 8 service units) the value of the objective function decreases very modestly (dropping by 0.18\% overall), while for larger number of steps the drops are much greater. We therefore decide to stop the decimation after 8 steps. This allows to switch-off 16\% of the service units (i.e. to save 16\% of the electric power) without affecting at all the service level.

\section{Conclusion\label{sec:conclusions}}
In this paper, we presented a novel computationally efficient  optimization approach  for
distributed resource allocation problems under user behavior uncertainty.  We propose a belief propagation scheme
to compute the costs of different service configurations. This is obtained by  averaging over all the possible
Nash equilibrium points associated to a given system configuration.

\section{Acknowledgements}
The authors acknowledge the european grants FET Open 265496, ERC 267915 and Italian FIRB Project RBFR10QUW4.


\begin{thebibliography}{99}

\small
\bibitem{V07}
B. Vocking. Selfish load balancing. In {\em Algorithmic game theory}. Cambridge Univ. Press, N. Nisan et al. Eds. (2007).

\bibitem{R73}
R. W. Rosenthal. A class of games possessing pure-strategy Nash equilibria. {\em Int. Journal of Game Theory}, {\bf 2}, 65-67 (1973)

\bibitem{R73b}
I. Milchtaich. Congestion games with player-specific payoff function. {\em Games and Economic Behavior}, {\bf 13}, 111-124 (1996).

\bibitem{KP99}
E. Koutsoupias and C. H. Papadimitriou. Worst-case equilibria.
{\em Symp. on Theoretical Aspects of Computer Science} (1999).

\bibitem{EDKM03}
E. Even-Dar, A. Kesselman, and Y. Mansour. Convergence time to Nash equilibria. In {\em Proc. 30th International Colloq. on Automata, Languages and Programming}, pp. 502-513 (2003).

\bibitem{FKKMS02}
D. Fotakis, et al.\ 
The structure and complexity of Nash equilibria for a selfish routing game. 29th ICALP, 123-134 (2002).


\bibitem{GI99}
A. Goel and P. Indyk. Stochastic Load Balancing and Related Problems.
{\em Symp.\ on Foundations of Computer Science} (1999).

\bibitem{KRT97}
 J. Kleinberg, Y. Rabani, and E. Tardos. Allocating bandwidth for bursty connections. {\em Proc. 29th ACM Symposium on Theory of Computing} (1997).

\bibitem{NSM11}
E. Nikolova and N. E. Stier-Moses. Stochastic selfish routing. In SAGT (2011).

\bibitem{DST03}
S. Dye, L. Stougie, and A. Tomasgard. The stochastic single resource service-provision problem. {\em Naval Research Logistics} {\bf 50}(8), 869-887 (2003).

\bibitem{ABRZ11}
F. Altarelli, A. Braunstein, A. Ramezanpour, and R. Zecchina, Stochastic Matching Problem, {\em Phys. Rev. Lett.} {\bf 106} (2011) 

\bibitem{MM09}
M. M\'ezard and A. Montanari {\em Information, Physics and Computation}. Oxford graduate texts (2009).


\bibitem{tech-rep}   F. Altarelli, A. Braunstein, L. Dall'Asta and R. Zecchina, {\em in preparation} (2013).

\end{thebibliography}
\end{document}